\documentclass[sn-mathphys-num]{sn-jnl}

\usepackage{graphicx}%
\usepackage{multirow}%
\usepackage{amsmath,amssymb,amsfonts}%
\usepackage{amsthm}%
\usepackage{mathrsfs}%
\usepackage[title]{appendix}%
\usepackage{xcolor}%
\usepackage{textcomp}%
\usepackage{manyfoot}%
\usepackage{booktabs}%
\usepackage{algorithm}%
\usepackage{algorithmicx}%
\usepackage{algpseudocode}%
\usepackage{listings}%

\usepackage{hyperref}
\hypersetup{
  colorlinks,%
  citecolor=black,%
  filecolor=black,%
  linkcolor=black,%
  urlcolor=blue
  }

\usepackage{orcidlink}
\newcommand{\orcidGP}{\orcidlink{0000-0003-2513-2459}} 
\newcommand{\orcidLS}{\orcidlink{0000-0003-4749-5250}}

\begin{document}

\title[The impact of AI]{The impact of artificial intelligence: from cognitive costs to global inequality}

\author[1]{\fnm{Guy} \sur{Pai\'c}\orcidGP{}}
\email{guy.paic@cern.ch} 

\author*[2]{\fnm{Leonid} \sur{Serkin}\orcidLS{}}
\email{lserkin@ciencias.unam.mx} 

\affil[1]{Instituto de Ciencias Nucleares, Universidad Nacional Autónoma de México,  Apartado Postal 70-543, Ciudad de México 04510, México}

\affil[2]{Facultad de Ciencias, Universidad Nacional Autónoma de México, Circuito Exterior s/n, Ciudad Universitaria, Coyoacán, Ciudad de México, 04510, México}

\abstract{In this paper, we examine the wide-ranging impact of artificial intelligence on society, focusing on its potential to both help and harm global equity, cognitive abilities, and economic stability. We argue that while artificial intelligence offers significant opportunities for progress in areas like healthcare, education, and scientific research, its rapid growth -- mainly driven by private companies -- may worsen global inequalities, increase dependence on automated systems for cognitive tasks, and disrupt established economic paradigms. We emphasize the critical need for strong governance and ethical guidelines to tackle these issues, urging the academic community to actively participate in creating policies that ensure the benefits of artificial intelligence are shared fairly and its risks are managed effectively.}

\keywords{artificial intelligence, impact, costs, risks, academic community}
\maketitle

\section{Introduction}

Artificial intelligence\footnote{For the purposes of this paper, we will adopt the latest definition of AI from the Organisation for Economic Co-operation and Development (OECD), which states~\cite{oecd2024principles}: “An AI system is a machine-based system that, for explicit or implicit objectives, infers, from the input it receives, how to generate outputs such as predictions, content, recommendations, or decisions that can influence physical or virtual environments. Different AI systems vary in their levels of autonomy and adaptiveness after deployment”.} (AI) is transforming the way we live. The impact of AI represents a rapid and transformative shift in society, comparable only to some of the most remarkable milestones in human history, such as the discovery of fire, the Industrial Revolution, or the invention of the automobile. Today, society is facing the rapid rise of AI, which -- like a massive tsunami -- is permeating every aspect of life. From sophisticated reinforcement learning algorithms that master chess and other games~\cite{silver} to AI-driven coding assistants~\cite{amin} and large language models (LLMs) like ChatGPT~\cite{ray} or DeepSeek~\cite{deepseek2024}, these innovations not only empower individuals to learn and innovate but also help build a more inclusive, interconnected global community.

\medskip
\noindent
However, every technological leap brings not only opportunities but also risks and unforeseen consequences. Human progress often comes with unintended side effects, such as climate change~\cite{Hansen} and plastic pollution~\cite{Jambeck}, and a significant portion of modern spending goes toward mitigating these risks inherent in our technologically interconnected society~\cite{wolff2021technology}. Our growing dependence on AI-driven systems -- whether in energy distribution, transportation, or healthcare -- can amplify the effects of any failure~\cite{Buldyrev}. In this context, the failure of any AI-driven decision-making process can trigger cascading disruptions similar to those observed in traditional infrastructural breakdowns~\cite{PrezCerrolaza2023, Danielsson}. These, and many other AI-driven developments illustrate that the transformative power of AI encompasses both remarkable opportunities and considerable challenges, making it essential to approach its governance and integration thoughtfully~\cite{Taeihagh2021, Zaidan2024}.

\medskip
\noindent
This paper examines the duality of AI’s impact -- its potential benefits versus its risks. We aim to focus on the less-discussed aspects -- specifically, the short- and long-term effects AI could have on humanity. We first examine the unintended consequences of technology, drawing lessons from historical advancements. Next, we discuss the need for governance and regulation, followed by an in-depth analysis of AI’s dangers, particularly in job markets and economic inequality. We then explore AI’s impact on healthcare, addressing both its benefits and ethical risks, before discussing the cognitive costs of AI, including dependency and skill erosion. Finally, we conclude by emphasizing the role of the academic community in shaping AI’s future.

\section{Collateral damage and effective governance}

Nowhere is the balance between technological progress and unintended harm more evident than in the widespread use of automobiles. While cars have revolutionized transportation and global connectivity, they also come with significant risks -- more than a million people lose their lives in traffic accidents each year~\cite{who2021traffic}. However, this risk is not uniform across different regions of the world. As seen in Table~\ref{tab:road_traffic}, geographic areas with stronger regulatory frameworks, such as vehicle safety standards, speed limits, and well-maintained infrastructure, report significantly lower fatality rates compared to those with weaker enforcement mechanisms~\cite{Tavakkoli2022}.

\medskip
\noindent
Beyond immediate fatalities caused by car accidents, the long-term consequences of increased car usage extend to rising obesity rates and elevated risks of cardiovascular disease and mortality~\cite{warren2010sedentary, sugiyama2020car}. This so-called collateral damage, however, is not an unavoidable consequence of technological advancement; rather, it is a governance challenge requiring effective policy intervention to mitigate harm~\cite{Peden2005}.

\medskip
\noindent
A similar governance approach is necessary for AI -- without clear oversight, AI could introduce new vulnerabilities, such as algorithmic bias, security threats, and unintended social consequences. As seen in the case of road safety, regions that implement strong regulations experience fewer negative consequences, reinforcing the need for proactive AI governance to ensure that technological progress benefits society while minimizing risks~\cite{Mikalef2022}.

\begin{table}[t]
\centering
\begin{tabular}{@{}lcc@{}} 
\toprule
\textbf{Region} & \textbf{Estimated road traffic death rate} & \textbf{Estimated number}\\
                & \textbf{(per 100,000 population)}         &  \textbf{of road traffic deaths}  \\
\midrule
Global          & 16.7 & 1,282,150 \\
Africa          & 27.2 & 297,087 \\
Eastern Mediterranean & 17.8 & 126,958 \\
Western Pacific & 16.4 & 317,393 \\
Southeast Asia  & 15.8 & 317,069 \\
Americas        & 15.3 & 154,780 \\
Europe          & 7.4  & 68,863 \\
\bottomrule
\end{tabular}
\caption{Estimated road traffic deaths data in 2019 by region, data taken from \cite{who2021traffic}.}
\label{tab:road_traffic}
\end{table}

\medskip
\noindent
This idea goes hand in hand with the growing movement within civil society to highlight the various risks associated with AI and to urge policymakers to address these concerns. The goal is to ensure that these risks are mitigated with the guidance of both the scientific community and the public~\cite{Alalawi2024}.

\medskip
\noindent
Historically, many groundbreaking innovations were the result of collaborations among public institutions, universities, and state-sponsored research initiatives~\cite{Etzkowitz2000}. However, in recent decades, especially in the field of digital technologies and AI, private companies have emerged as the primary drivers of transformative innovations, often operating with minimal direct societal oversight~\cite{Hagendorff2023, Lundvall}.

\medskip
\noindent
This shift contrasts with the approach taken by the European Organization for Nuclear Research (CERN) regarding the World Wide Web. In 1993, rather than patent or privatise the web’s source code, CERN released it freely to the public~\cite{cern2024web}. This act ensured that the web would remain an open platform for global innovation and collaboration, free from proprietary restrictions. By adopting this open approach, CERN enabled the explosive growth of the internet, creating countless opportunities for businesses, education, and communication across the globe, a legacy that contrasts with today's more closed, profit-driven models of technological development.

\medskip
\noindent
Given AI's profound societal impact, adopting a similar multinational, nonprofit-driven approach to its development could help ensure its benefits are equitably shared. Promoting global collaboration within an open framework -- rather than leaving AI’s trajectory solely in the hands of private interests -- could lead to more ethical, transparent, and broadly beneficial technological advancements.

\section{AI’s dual impact}

There is no doubt that AI offers significant benefits. As the OECD states, "AI holds the potential to address complex challenges, from enhancing education and improving healthcare to driving scientific innovation and climate action"~\cite{oecd2024ai}. However, the risks associated with AI should not be underestimated.

\medskip
\noindent
The existential risks posed by AI, particularly the loss of millions of jobs, have been highlighted by various experts -- around 40\% of all working hours could be impacted by AI LLMs such as ChatGPT-4~\cite{Eloundou2024}. If not properly controlled, AI could widen existing inequalities and reshape entire industries, potentially leaving many workers without meaningful employment ~\cite{Frey2017, Shen2024}. This situation brings to mind an anecdote of a United Nations expert observing a peasant ploughing his field with a donkey: “We will give you a tractor to plough your piece of land in two hours instead of you ploughing your field in a whole day.” The peasant's reply was quick: “Well, what would my donkey and I do for the rest of the day?”

\medskip
\noindent
The potential job losses due to AI, even with compensation, have not been fully addressed by governments, and the psychological impact could be significant. As societies become increasingly dependent on AI-driven systems and digital communication, we are already witnessing broader social changes, particularly in rural communities~\cite{Correa2016}. Economic shifts and urban migration have contributed to the decline of traditional social spaces, such as village cafés, which once served as key hubs of local interaction. While modern communication tools provide new ways to stay connected, they do not fully replace in-person social interactions, which remain essential for community cohesion and mental well-being~\cite{Ettman2023}.

\medskip
\noindent
Beyond its impact on labour markets and social structures, AI is also reshaping critical sectors such as healthcare. There is clear evidence that AI holds immense potential to revolutionise medical diagnostics, enhance treatment strategies, and support healthcare professionals in delivering more precise and efficient care. However, despite these advancements, concerns persist regarding the ethical implications and unintended consequences of AI-driven healthcare tools. The World Health Organization (WHO) urges caution in the use of AI tools, particularly LLMs, to ensure they promote human well-being, safety, and autonomy while safeguarding public health~\cite{who2023aihealth}.

\medskip
\noindent
One of the most concerning dangers in using AI-driven innovations is its potential to worsen racial, gender, and geographic disparities in healthcare. This is because bias is often embedded in the data used to train AI models, which can lead to unequal treatment and outcomes for different groups of people \cite{turchin2020risks}. This presents an additional challenge for less developed countries, which must ensure the collection, privacy, and secure storage of large, representative datasets \cite{cenia2023index}.

\medskip
\noindent
Currently, WHO supports the responsible use of AI to benefit healthcare professionals, patients, and researchers. However, they emphasise the need for ethical guidelines and appropriate governance, as outlined in the WHO's guidance on AI ethics in healthcare \cite{who2021aiethics}. This also reinforces our perspective: the same level of careful scrutiny applied to other new technologies must also be consistently applied to LLMs.

\section{Cognitive cost}

The application of AI and its derivatives requires both human and industrial resources. Just as with traffic-related hidden damages mentioned earlier, there are several important aspects of AI development that we would like to address.

\medskip
\noindent
First, there is the risk of widening the “digital divide” between the most developed nations and those that are moderately developed or entirely disadvantaged. AI’s immense power and water demands are already proving to be a challenge, even for advanced economies~\cite{Luccioni2023, USAIAct2024}. According to studies from the International Monetary Fund (IMF), the computational power required to sustain AI development is growing rapidly, with the potential to consume as much energy as entire countries in the near future \cite{imf2024aiwork}. 

\medskip
\noindent
This raises critical questions about whether smaller countries, lacking the necessary infrastructure to support such large power demands, will ever be able to participate in AI development at that scale~\cite{Armstrong2016}. The real concern is whether AI will become a powerful tool controlled by a small number of countries, much like nuclear weapons or rocket technology~\cite{Schmid2025}. The IMF's conclusion is clear: “Emerging market and developing economies should prioritize developing digital infrastructure and digital skills”~\cite{imf2024aiwork}.

\medskip
\noindent
To prevent technological competition from leading to unnecessary environmental sacrifices, there is an urgent need for collaborative governance that establishes binding international standards. Cooperation between governments and technology companies can enable the sustainable development of AI, ensuring that climate goals are protected without suppressing innovation~\cite{Francisco2023}.

\medskip
\noindent
On the positive side, empirical and historical analyses indicate that, although many technological breakthroughs originate in advanced economies, such innovations have often acted as catalysts for accelerated socio-economic convergence in developing regions~-- a phenomenon extensively documented in studies on technological diffusion and socio-economic progress~\cite{pinker}.

\medskip
\noindent
Second, we must consider the long-term effects of AI on human cognitive abilities. While AI is intended to relieve humanity of many mental tasks, it is unclear whether this will be a benefit in the long run. Younger generations are already shifting their reliance on cognitive skills. Tasks like memorising phone numbers, solving maths problems, or even learning new languages are becoming obsolete with the rise of mobile phones and AI-powered translators \cite{leon2024cognitive, shanmugasundaram2023impact}.

\medskip
\noindent
Finally, what happens if, for some reason, access to AI systems is lost? At present, people still possess the skills to revert to pre-AI methods, much like pilots who are instructed to override AI systems if they do not understand its actions \cite{easa2024ai}. However, as we mentioned earlier, excessive reliance on AI could gradually erode human cognitive skills~\cite{Ahmad2023, Psychol}. To ensure resilience, it is crucial to preserve our ability to think critically, adapt, and function independently of AI -- both in everyday life and in times of crisis.

\section{Conclusions: our role as the academic community}

The academic community plays a crucial role in shaping the development and responsible oversight of AI, guiding its future use in ways that benefit society while addressing its risks. Universities and research institutions are at the forefront of AI development, carrying the unique responsibility of applying AI in a controlled and informed manner. 

\medskip
\noindent
To achieve this, academia should lead the way in exploring potential risks posed by AI, such as job displacement, privacy concerns, health, and ethical challenges. One of academia’s key roles is sharing knowledge, particularly about the dangers and potential misuse of AI. With their technical expertise, academics are well-positioned to provide guidance on the legislative and political actions needed to regulate AI effectively~\cite{ties}.

\medskip
\noindent
Finally, since AI is a global phenomenon, academic institutions worldwide should collaborate to establish international standards for its governance, helping to ensure AI does not deepen inequalities or contribute to geopolitical tensions. By implementing these strategic measures, the academic community can actively guide AI development to ensure it remains both innovative and ethically responsible.

\section{Acknowledgments}
The research was supported by the Mexican National Council of Humanities, Sciences and Technologies CONAHCYT under Grants No. CF-2042 and No. A1-S-22917. The authors are grateful to Gergely G\'abor Barnaf\"oldi for his valuable comments, and sincerely appreciate the valuable feedback from the journal reviewers.

\section{Data availability statement}
The dataset on the estimated road traffic deaths data in 2019 is available at: \url{https://apps.who.int/gho/data/node.main.A997}.


\end{document}